\documentclass[prl,aps,twocolumn,showpacs,floatfix]{revtex4-1}
\usepackage{amsfonts}
\usepackage{amssymb}
\usepackage{hyperref}
\usepackage{graphicx}
\usepackage{dcolumn}
\usepackage{bm,amsmath,verbatim}
\usepackage{mathrsfs}
\usepackage{color}

\hypersetup{colorlinks,
linkcolor=blue,          citecolor=blue,        filecolor=blue,      urlcolor=blue           }

\def\be{\begin{equation}}
\def\ee{\end{equation}}
\def\bea{\begin{eqnarray}}
\def\eea{\end{eqnarray}}
\def\bse{\begin{subequations}}
\def\ese{\end{subequations}}

\def\be{\begin{eqnarray}}
\def\ee{\end{eqnarray}}

\begin{document}

\title{Weyl Exceptional Rings in a Three-Dimensional Dissipative Cold Atomic
Gas}
\author{Yong Xu}
\email{yongxuph@umich.edu}
\author{Sheng-Tao Wang}
\author{L.-M. Duan}
\affiliation{Department of Physics, University of Michigan, Ann Arbor,
Michigan 48109, USA}

\begin{abstract}
Three-dimensional topological Weyl semimetals can generally support a zero-dimensional
Weyl point characterized by a quantized Chern number or a
one-dimensional Weyl nodal ring (or line) characterized by a quantized Berry phase
in the momentum space.
Here, in a dissipative system with particle gain and loss, we discover
a new type of topological ring, dubbed Weyl exceptional ring consisting of
exceptional points at which two eigenstates coalesce. Such a Weyl exceptional
ring is characterized by both a quantized Chern number and a quantized Berry phase,
which are defined via the Riemann surface. We propose an experimental scheme to
realize and measure the Weyl exceptional ring in a dissipative cold atomic gas trapped
in an optical lattice.
\end{abstract}

\maketitle
Recently, condensed matter systems have proven to be a powerful platform
to study low energy gapless particles by using momentum space band structures to mimic the
energy-momentum relation of relativistic particles~\cite{Burkov2016,Jia2016} and
beyond~\cite{Burkov2011PRB,Yong2015PRL,Soluyanov2015Nature,Bernevig2016Science}.
One celebrated example in three dimensions is the zero-dimensional Weyl
point~\cite{Wan2011prb,Yuanming2011PRB,Burkov2011PRL,ZhongFang2011prl,LingLu2013NP,
Gong2011prl,Anderson2012PRL,Yong2014PRL,Shengyuan2014PRL,Tena2015RPL,Bo2015,Ueda2016} described
by the Weyl Hamiltonian, which has been long sought-after in particle physics but only experimentally
observed in condensed matter materials~\cite{Lu2015,Xu2015,Lv2015}. Such a Weyl point
can be viewed as a magnetic monopole~\cite{volovik} in the momentum space
and possesses a quantized Chern number on a surface enclosing the point.
Another example is the one-dimensional Weyl nodal ring~\cite{Burkov2011PRB,HasanRing,Yong2016PRA,DWZhang2016},
which has no counterpart in particle physics. It can be regarded as the generalization of zero-dimensional
Dirac cones in two-dimensional systems, such as in graphene, to three-dimensional systems.
Such a nodal ring has a quantized Berry phase over a closed path encircling it
but does not possess a nonzero quantized Chern number. This leads to a natural question of whether
there exists a topological ring exhibiting both a quantized Chern number and a quantized
Berry phase in the momentum space.

So far, studies on those gapless states focus on closed and lossless systems.
However, particle gain and loss are generally present in natural systems.
Such systems can often be described by non-Hermitian Hamiltonians~\cite%
{Moiseyev2011, Berry2004, Rotter2009, Heiss2012}, which are widely applied
to many different systems~\cite{Guo2009PRL, Kip2010,
Stone2011,Yang2014,Moljacic2015,
Zoller2012,LuoleArxiv,Liertzer2012PRL,
Zhang2014Science, Khajavikhan2014Science,Gao2015}. Due to the
non-Hermiticity, eigenvalues of the Hamiltonian are generically complex
unless the $\mathcal{PT}$ symmetry~\cite{Bender1998} is conserved and the
imaginary part of energy is associated with either decay or growth.
Another intriguing feature of a non-Hermitian system is the existence of
exceptional points (EPs)~\cite{Moiseyev2011, Rotter2009, Berry2004,
Heiss2012} at which two eigenstates coalesce and the Hamiltonian becomes
defective, leading to many novel phenomena, such as loss-induced
transparency~\cite{Guo2009PRL}, single-mode lasers~\cite%
{Khajavikhan2014Science,Zhang2014Science}, and reversed pump dependence of
lasers~\cite{Liertzer2012PRL}.

In this paper, we investigate a system of Weyl points in the presence of a
spin-dependent non-Hermitian term and find a Weyl exceptional ring
composed of EPs. In stark contrast to a Weyl
nodal ring~\cite{Burkov2011PRB,Yong2016PRA,DWZhang2016}, which does not have
a nonzero Chern number, remarkably, this ring exhibits a nonzero quantized
Chern number as long as the integral of the Berry curvature is evaluated
over a surface (labeled by $\mathcal{S}$) that encloses the whole ring.
Since energy is multi-valued in the complex parameter space due to its
square root form, a state on the surface $\mathcal{S}$ may be defined over
the Riemann surface, on which a function is single valued. On the other
hand, the Chern number is zero when the surface $\mathcal{S}$ does not
enclose any part of the ring even when it is located inside it. Besides the
Chern number, such a Weyl exceptional ring has a quantized Berry phase over
a trajectory encircling the ring twice, instead of once in the case of the
Weyl nodal ring. Furthermore, we propose a feasible scheme to engineer and
probe the Weyl exceptional ring in a dissipative ultracold atomic gas. In
such a system, we find that the Fermi arc can still exist but is suppressed,
even though the Weyl point transforms into a ring.

\begin{figure*}[t]
\includegraphics[width=\textwidth]{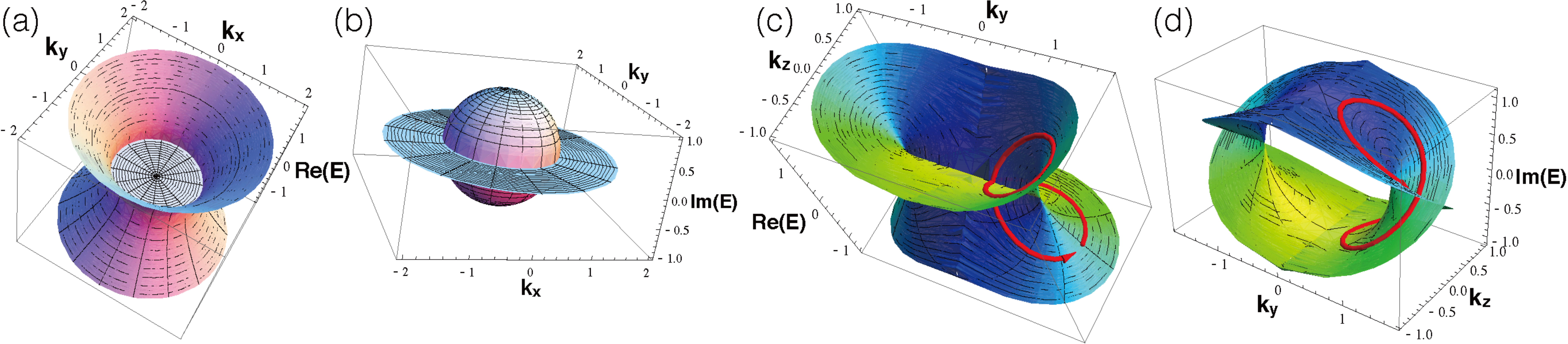}
\caption{(Color online) Energy spectra and the Riemann surface of the toy
model in Eq.~\eqref{Eq:toy}. Spectra with respect to $k_x$ and $k_y$ for $%
k_z=0$ in (a) (real parts) and (b) (imaginary parts). Real (c) and imaginary
parts (d) of the Riemann surface as a function of $k_y$ and $k_z$ for $k_x=0$%
. In (c) and (d), the color represents the strength of $\protect\theta \text{
mod } 4\protect\pi$, and the red tube arrow shows a path from $\protect\theta%
=0$ to $\protect\theta=4\protect\pi$.}
\label{Fig1}
\end{figure*}

\emph{Toy model of Weyl exceptional ring}---Near a Weyl point in the
momentum space, a system can be described by the Weyl Hamiltonian $H_{W}=\pm
\sum_{\nu }v_{\nu }k_{\nu }\sigma _{\nu }$, where $\sigma _{\nu }$ represent
Pauli matrices and $\pm $ the chirality. For clarity, we consider the
positive chirality and choose $v_{\nu }=1$ hereafter. In the presence of a
non-Hermitian term $i\gamma \sigma _{z}$ ($\gamma >0$) associated with
particle gain for spin up atoms and loss for spin down ones, the Hamiltonian
becomes
\begin{equation}
H(\mathbf{k})=\sum_{\nu =x,y,z}k_{\nu }\sigma _{\nu }+i\gamma \sigma _{z},
\label{Eq:toy}
\end{equation}%
taking the energy unit to be $1$. The eigenvalues are $E_{\theta }(\mathbf{k}%
)=\sqrt{k^{2}-\gamma ^{2}+2ik_{z}\gamma }=\sqrt{A(\mathbf{k})}e^{i\theta /2}$%
, where $A(\mathbf{k})=\sqrt{(k^{2}-\gamma ^{2})^{2}+4k_{z}^{2}\gamma ^{2}}$
with $k^{2}=k_{x}^{2}+k_{y}^{2}+k_{z}^{2}$, and $\theta $ is defined via $%
\cos \theta =(k^{2}-\gamma ^{2})/A(\mathbf{k})$ and $\sin \theta
=2k_{z}\gamma /A(\mathbf{k})$. Here, $\theta $ is used to label two
branches, given that $e^{i\theta /2}$ gains a minus sign upon $\theta
\rightarrow \theta +2\pi $, corresponding to the other band. In the absence
of $\gamma $, energy of both bands is real and two bands touch at $%
\mathbf{k}=0$ with linear dispersion along all three momentum directions. In
this case, $\theta $ takes only two nonequivalent discrete values: $0$ and $%
2\pi $ (corresponding to two distinct and separate bands). When $\gamma >0$,
the eigenvalues become complex, and the single touching point morphs into a
Weyl exceptional ring in the $k_{z}=0$ plane characterized by $%
k_{x}^{2}+k_{y}^{2}=\gamma ^{2}$. On this ring, both the real and imaginary
parts of the eigenvalues vanish [shown in Fig.~\ref{Fig1}(a) and (b)] and
two eigenstates coalesce into a single one (different from the case of
degeneracy). Fig.~\ref{Fig1}(a) and (b) also illustrate that in the $k_{z}=0$
plane energy is purely real outside the ring (with conserved $\mathcal{PT%
}$ symmetry) and purely imaginary inside it for this simple model.
Interestingly, $\theta $ takes continuous values from $0$ to $4\pi $ ($%
\theta $ and $\theta +4\pi $ are equivalent) and gains $2\pi $ when a state
travels through the ring and returns, ending up at another state with
opposite energy, arising from the role of branch points that the Weyl
exceptional ring plays.

In complex analysis, besides using branch cuts, an alternative visual
representation to depict a multi-valued function is the Riemann surface, a
two-dimensional (2D) manifold that wraps around the complex plane infinite
(noncompact) or finite (compact) number of times. Before we discuss the
topology of the Weyl exceptional ring, let us first focus on the definition
of a closed 2D surface $\mathcal{S}$ via the Riemann surface. In Fig.~\ref%
{Fig1}(c) and (d), we plot the Riemann surface of $E_{\theta }$ for $k_{x}=0$
(the color represents the strength of $\theta \text{ mod }4\pi $), showing
that energy is single-valued on the surface, which connects the
different bands. Given the single value property, we define each state on $%
\mathcal{S}$ to be living on the Riemann surface. For example, if we
consider a state at $\mathbf{k}_{0}$ with $\theta _{0}$, any other states on
the surface $\mathcal{S}$ can be obtained by starting from this state and
travelling on the momentum space surface $\mathcal{S}$ while keeping $%
E_{\theta }(\mathbf{k)}$ on the Riemann surface.

\begin{figure}[t]
\includegraphics[width=3.2in]{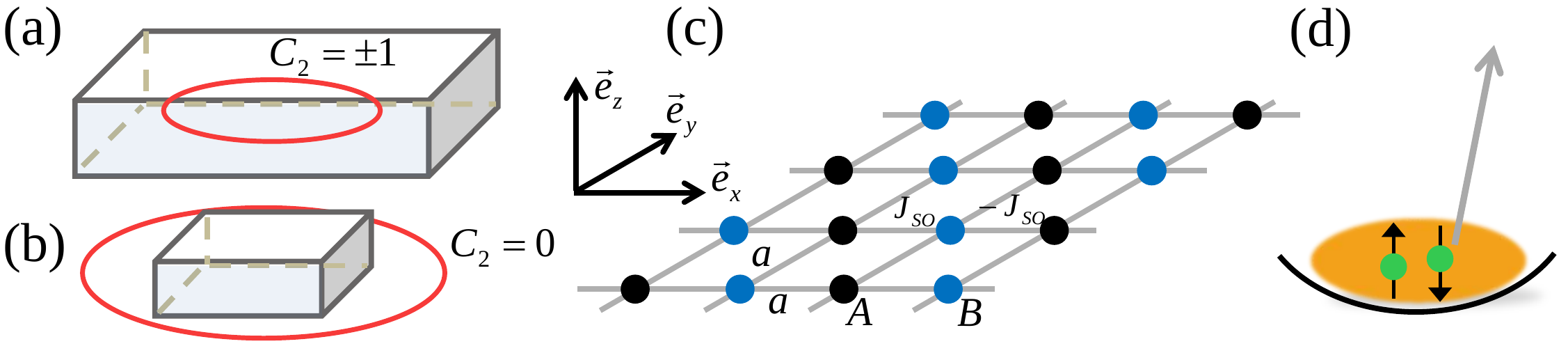}
\caption{(Color online) (a) A surface enclosing a Weyl exceptional ring and
(b) a surface located inside the ring. (c) Lattice structure in the $(x,y)$
plane. (d) Schematic of trapped atoms being kicked out by a resonant optical
beam (denoted by the grey arrow).}
\label{Fig2}
\end{figure}

With the proper definition of a closed 2D surface, we can characterize the
topology of a Weyl exceptional ring by the Chern number on the surface based
on two approaches: the integral of spin vector fields and the Berry
curvature. For the former, the Chern number is given by~\cite{KlemanSoft}
\begin{equation}
N_3=\frac{1}{4\pi}\oint_{\mathcal{S}}\mathbf{d}_\theta \cdot \left(\frac{%
\partial \mathbf{d}_\theta}{\partial u_1}\times\frac{\partial \mathbf{d}%
_\theta} {\partial u_2}\right)du_1 du_2,
\end{equation}
which characterizes the number of times that the spin field $\mathbf{d}%
_\theta=\sum_{\nu=x,y,z}\langle\sigma_\nu\rangle \mathbf{e}_\nu$ wraps
around a closed surface $\mathcal{S}$ parametrized by ($u_1$,$u_2$). Here, $%
\mathbf{e}_\nu$ denotes the unit vector along the $\nu$ direction and $%
\langle\sigma_\nu\rangle\equiv \langle u_{\theta}(\bm k)|\sigma_\nu|u_{%
\theta}(\bm k)\rangle$ with $|u_{\theta}(\bm k)\rangle$ being the normalized
right eigenstate of $H(\mathbf{k})$ [i.e., $H(\mathbf{k})|u_{\theta}(\bm %
k)\rangle=E_\theta(\mathbf{k})|u_{\theta}(\bm k)\rangle$ and $\langle
u_{\theta}(\bm k)|u_{\theta}(\bm k)\rangle=1$]. Direct calculations show
that $N_3=\pm 1$ when the surface $\mathcal{S}$ encloses the whole ring as
shown in Fig.~\ref{Fig2}(a), while $N_3=0$ when it does not enclose any part
of the ring [shown in Fig.~\ref{Fig2}(b)].

Analogous to the scenario without decay~\cite{XiaoRMP}, we may also define
the first Chern number via the Berry curvature
\begin{equation}
C_2=\frac{1}{2\pi}\oint_{\mathcal{S}}{\bm \Omega_\theta}(\bm k)\cdot d{\bm S}%
,  \label{ChernEq}
\end{equation}
where ${\bm \Omega}_{\theta}(\bm k)=i\langle \nabla_{\bm k} u_{\theta}(\bm %
k)|\times |\nabla_{\bm k} u_{\theta}(\bm k)\rangle$ is the Berry curvature.
Our calculations show that $C_2=\pm 1$ when the surface $\mathcal{S}$
encloses the Weyl exceptional ring and $C_2=0$ otherwise, suggesting that
the topological charge is entirely carried by the ring.

The physical meaning of the Berry curvature in this system can be understood
from the following semiclassical equation under an external gradient force $%
\mathbf{F}$ (see the supplemental material for derivation)
\begin{eqnarray}
\dot {\mathbf{r}}_{c}&=&\partial_{\mathbf{k}_{c}}\bar{E}({\mathbf{k}_{c}})-%
\dot {\mathbf{k}}_{c}\times{\bm \Omega_\theta}({\bm k}_c), \\
\hbar\dot {\mathbf{k}}_{c}&=&\mathbf{F},
\end{eqnarray}
where $\bar{E}({\mathbf{k}_{c}})=\text{Re}[E_\theta({\mathbf{k}_{c}})]+\bar{%
\mathbf{A}}_\theta(\mathbf{k}_{c}) \cdot {\dot {\mathbf{k}}}_{c}$, $\bar{%
\mathbf{A}}_\theta(\mathbf{k}_{c})\equiv\text{Re}[\mathbf{A}_\theta(\mathbf{k%
}_{c})-\tilde{\mathbf{A}}_\theta(\mathbf{k}_{c})]$ with the Berry connection
being $\mathbf{A}_\theta(\mathbf{k})=i\langle{u}_\theta(\mathbf{k}%
)|\partial_{\mathbf{k}}u_\theta(\mathbf{k})\rangle$ and $\tilde{\mathbf{A}}%
_\theta(\mathbf{k})=i\langle\tilde{u}_\theta(\mathbf{k})|\partial_{\mathbf{k}%
}u_\theta(\mathbf{k})\rangle$, where $\langle \tilde{u}_{\theta}(\bm k)|$ is
the normalized left eigenstate of $H$ [i.e., $\langle \tilde{u}_{\theta}(\bm %
k)|H(\mathbf{k})=\langle \tilde{u}_{\theta}(\bm k)|E_\theta(\mathbf{k})$ and
$\langle\tilde{u}_\theta(\mathbf{k})|u_\theta(\mathbf{k})\rangle=1$]; $%
\mathbf{r}_c$ and $\mathbf{k}_c$ are the center coordinate of a wave packet
in the real space and momentum space, respectively. Clearly, the Berry
curvature plays the same role as in the traditional semiclassical equation
in a closed system~\cite{Niu1999}. However, in this open system, the
equation includes a term that effectively modifies the energy spectra,
resulting from the difference between left and right eigenstates, a feature
in a non-Hermitian Hamiltonian. Without $\mathbf{F}$, the group velocity is
dictated by the real part of the spectra, which implies that inside the Weyl
exceptional ring in the $k_z=0$ plane, the group velocity vanishes.

Other than the Chern number on the surface, there also exists a quantized
Berry phase characterizing the Weyl exceptional ring, defined as
\begin{equation}
C_1=\oint_{2\mathcal{L}}{i\langle\tilde{u}_\theta(\mathbf{k})|\partial_{%
\mathbf{k}} u_\theta(\mathbf{k})\rangle} \cdot d{\bm k},  \label{BerryEq}
\end{equation}
where the path $2\mathcal{L}$ travels across the ring twice along the
Riemann surface so that the state returns to the original one after the
entire trajectory as shown in Fig.~\ref{Fig1}(c) and (d). Direct
calculations yield $C_1=\pm\pi$, consistent with the result for a single EP~%
\cite{Moiseyev2011}. This Berry phase is different from that of a Weyl nodal
ring in which the quantized Berry phase is obtained when the trajectory
encircles the ring once~\cite{Burkov2011PRB,Yong2016PRA,DWZhang2016}.

\emph{Realization in dissipative cold atomic gases}---To realize the Weyl
exceptional ring in cold atoms, we consider the following model
\begin{eqnarray}
H = && \sum_{k_{z},\mathbf{x}} \Big[ (\bar{h}_{z}+i\gamma)\hat{c}_{k_{z},%
\mathbf{x}}^{\dagger }\sigma_z\hat{c}_{k_{z},\mathbf{x}} + \sum_{\nu=x,y} (-J%
\hat{c}_{k_{z},\mathbf{x}}^{\dagger }\hat{c}_{k_{z},\mathbf{x}+ a\mathbf{e}%
_\nu}  \notag \\
&& +(-1)^{j_x+j_y}J_{SO\nu}\hat{c}_{k_{z},\mathbf{x}}^{\dagger }\sigma_\nu%
\hat{c}_{k_{z},\mathbf{x}+ a\mathbf{e}_\nu }+H.c. )+h_0 \Big],
\end{eqnarray}
where $\mathbf{x}=j_{x}a\mathbf{e}_x+j_{y}a\mathbf{e}_y$ (with $a$ being the
lattice constant) labels the location of sites, $\hat{c}^\dagger_{k_{z},%
\mathbf{x}}=(
\begin{array}{cc}
\hat{c}^\dagger_{k_{z},\mathbf{x},\uparrow} & \hat{c}^\dagger_{k_{z},\mathbf{%
x},\downarrow}%
\end{array}
)$ with $\hat{c}_{k_{z},\mathbf{x},\sigma }^{\dagger }$ ($\hat{c}_{k_{z},%
\mathbf{x},\sigma }$) being the creation (annihilation) operator, $J$ and $%
J_{SO\nu}$ ($J_{SOx}=-J_{SOy}=J_{SO}$) stand for the tunneling and
spin-orbit coupling strength, $h_0=[-i\gamma+\hbar ^{2}k_{z}^{2}/(2m)]\hat{c}%
_{k_{z},\mathbf{x}}^{\dagger }\hat{c}_{k_{z},\mathbf{x}}$, with $\gamma$
denoting the decay strength, and $\bar{h}_{z}=\alpha k_{z}+h_{z}$ is the
effective Zeeman field with $\alpha=\hbar^2 k_{Lz}/(2m)$ where $k_{Lz}$
depends on the wave vector of Raman laser beams along the $z$ direction, $m$
is the mass of atoms, and $h_{z}$ the Zeeman field proportional to the
two-photon detuning. Here, we consider the atoms to be trapped in an optical
lattice in the $x$ and $y$ directions while there is no lattice along the $z$
direction.

Without $\gamma$, this Hamiltonian, which has two Weyl points and a fourfold
degenerate point, can be experimentally engineered by coupling two hyperfine
states with two pairs of Raman laser beams in cold atom optical lattices~%
\cite{Yong2016Arxiv}. To generate the decay term representing an atom loss $%
-2i\gamma$ for spin down atoms, one may consider using a resonant optical
beam to kick the atoms in the $|{\downarrow}\rangle$ state out of a weak
trap as shown in Fig.~\ref{Fig2}(d), which has been experimentally realized
in $^6$Li~\cite{LuoleArxiv}. Alternatively, one may consider applying a
radio frequency pulse to excite atoms in the $|{\downarrow}\rangle$ state to
another irrelevant state $|3\rangle$, leading to an effective decay for spin
down atoms when atoms in $|3\rangle$ experience a loss by applying an
anti-trap.

To see the energy spectra, we write down the Hamiltonian in the momentum
space,
\begin{equation}
H(\mathbf{k})=(\bar{h}_{z}+i\gamma
)\sigma_{z}-h_{t}\tau_{x}+\tau_{y}(-b_{x}\sigma_{x}+b_{y}\sigma_{y}),
\label{HTBk}
\end{equation}
in the basis $\Psi(\mathbf{k})^{T}$ with $\Psi(\mathbf{k})=(%
\begin{array}{cccc}
e^{ik_{x}a}\hat{A}_{\mathbf{k}\uparrow} & e^{ik_{x}a}\hat{A}_{\mathbf{k}%
\downarrow} & \hat{B}_{\mathbf{k}\uparrow} & \hat{B}_{\mathbf{k}\downarrow}%
\end{array}%
)$, where $\hat{A}_{\mathbf{k}\sigma}$ ($\hat{B}_{\mathbf{k}\sigma}$)
annihilates a state with spin $\sigma$ and momentum $\mathbf{k}$ located at $%
A$ ($B$) site [$A$ and $B$ constitute a unit cell as shown in Fig.~\ref{Fig2}%
(c)]. Here, $h_{t}=2J [\cos (k_{x}a) + \cos (k_{y}a) ]$, $b_{x}=2J_{SO}\sin
(k_{x}a)$ and $b_{y}=-2J_{SO}\sin(k_{y}a)$; $\tau_{x,y}$ are Pauli matrices
acting on $A,B$ sublattices. This Hamiltonian can be transformed into a block
diagonal matrix, i.e., $H\rightarrow H^\prime=(\bar{h}_{z}+i\gamma
)\sigma_{z}-h_{t}\sigma_z\tau_{z}+ \tau_{z}(b_{x}\sigma_{y}+b_{y}\sigma_{x})$%
, which commutes with $\tau_z$. Note that we have neglected the
spin-independent term $h_0$, which has no essential effects on the physics.

\begin{figure}[t]
\includegraphics[width=3.4in]{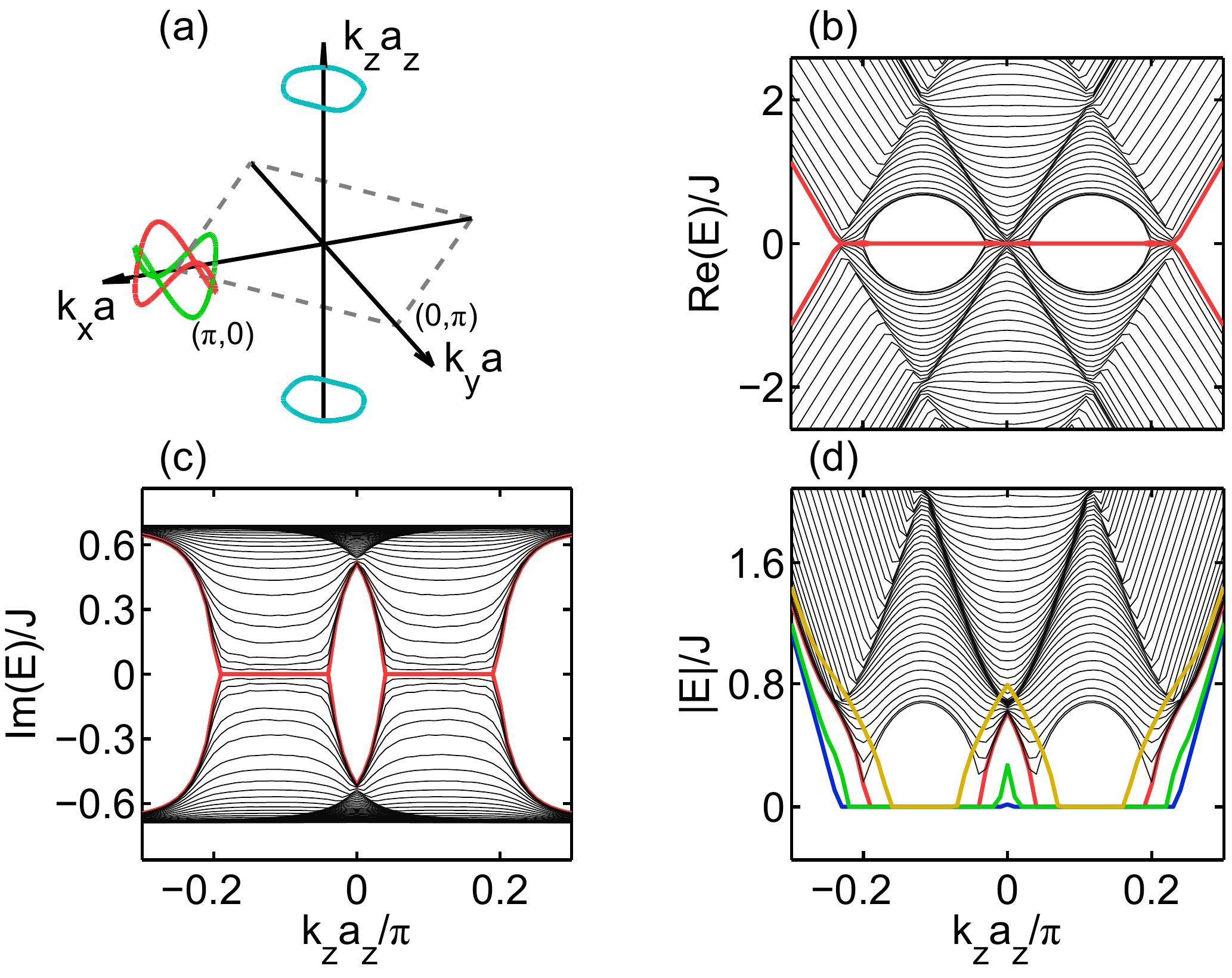}
\caption{(Color online) (a) Schematic of Weyl exceptional rings denoted by
closed red, green and cyan lines for the system described by the Hamiltonian
in Eq.~\eqref{HTBk}. The dashed box depicts the first Brillouin zone. (b)
Real (c) imaginary and (d) absolute values of the eigenenergy with respect
to $k_za_z$ for $k_y=0$ and $\protect\gamma=0.7J$ when the open boundary
condition is imposed along the $x$ direction. The red lines are the surface
states. In (d), additional surface states for $\protect\gamma=0, 0.35J, 0.86J
$ are plotted as blue, green and yellow lines respectively. Note that only
the parts with zero absolute energy are associated with surface states.}
\label{Fig3}
\end{figure}

Similar to the toy model in Eq.~\eqref{Eq:toy}, eigenvalues of this
Hamiltonian are $E_{\theta_\pm}(\mathbf{k})=\sqrt{b_{\pm}^2-\gamma^2+2ib_{z%
\pm}\gamma}=\sqrt{A_{\pm}(\mathbf{k})}e^{i\theta_{\pm}/2}$, where $A_{\pm}(%
\mathbf{k})=\sqrt{(b_{\pm}^2-\gamma^{2})^{2}+4b_{z\pm}^{2}\gamma^{2}}$ with $%
b_{\pm}^2=b_{x}^{2}+b_{y}^{2}+b_{z\pm}^{2}$ and $b_{z\pm}=\pm h_{t}+\bar{h}%
_{z}$ ($\pm$ label two particle or hole bands associated with the subspace $%
\tau_z=\mp$ for $H^\prime$), and $\theta_\pm$ are defined by $%
\cos\theta_{\pm}=(b_{\pm}^2-\gamma^{2})/A_{\pm}(\mathbf{k})$ and $%
\sin\theta_{\pm}=2b_{z\pm}\gamma/A_{\pm}(\mathbf{k})$. Without $\gamma$,
energy is purely real, and Weyl points emerge at $\mathbf{k}%
_{W0}=(k_{x}a,k_{y}a,k_z a_z)=[\pi,0,-2m\pi h_z/(\hbar^2 k_{Lz}^2)] $ or $%
\mathbf{k}_{W\pm}=[0,0,-2m\pi(\pm 4 J+h_z)/(\hbar^2 k_{Lz}^2)]$, where $%
a_z=\pi/k_{Lz}$. The touching point is fourfold (doubly) degenerate at $%
\mathbf{k}_{W0}$ ($\mathbf{k}_{W\pm}$). When $\gamma > 0$, the spectrum
becomes complex and it is purely real only in the plane $b_{z\pm}=0$. A
touching point transforms into a closed line (i.e., Weyl exceptional ring)
at which particle and hole bands coalesce when $b_{z\pm}=0$ and $%
b_x^2+b_y^2=\gamma^2$, as shown in Fig.~\ref{Fig3}(a). Around $\mathbf{k}%
_{W0}$, the fourfold degeneracy of the touching point is broken, and there
emerge two Weyl exceptional rings that are not degenerate except at four
points with $|\sin{k_x a_x}|=\gamma/(2\sqrt{2}J_{SO})$, $k_x=\pm k_y-\pi$,
and $k_z=-h_z/\alpha$ [as shown in Fig.~\ref{Fig3}(a)]. Around $\mathbf{k}%
_{W\pm}$, each Weyl point morphs into a single Weyl exceptional ring, which
can be approximated by $k_x^2+k_y^2=\gamma^2/(4J_{SO}^2)$ and $k_z=[-h_z\pm
J(4-\gamma^2/(4J_{SO}^2))]/\alpha$ when $\gamma\ll 2J_{SO}$.

Analogous to the toy model, a Weyl exceptional ring in this realistic model
can be characterized by the Chern number defined in Eq.~\eqref{ChernEq},
i.e., evaluated by an integral of the Berry curvature over a closed surface $%
\mathcal{S}$ via the Riemann surface. Around $k_{W0}$, there are two Weyl
exceptional rings associated with two branches $\theta_{\pm}$, and the Chern
number is defined for each band with $C_{\theta_\pm}=1$ ($%
C_{\theta_\pm+2\pi}=-1$) when $\mathcal{S}$ encloses one ring. Around $%
k_{W\pm}$, the corresponding band contributes $C_{\theta_\pm}=-1$ ($%
C_{\theta_\pm+2\pi}=1$). Furthermore, apart from the Chern number, we can
characterize the ring by a quantized Berry phase defined in Eq.~%
\eqref{BerryEq}, i.e., evaluated along a closed trajectory enclosing the
Weyl exceptional ring twice for a considered band with a ring.

Another intriguing feature of Weyl semimetals is the existence of a Fermi
arc, surface states that connect two Weyl points with opposite Chern numbers
in a geometry with edges. When $\gamma >0$, Weyl points develop into Weyl
exceptional rings, and one may wonder on the existence of surface states
with open boundaries. Here, we calculate the spectra of the open system
under open boundaries along the $x$ direction and plot the real, imaginary
and absolute parts of the spectra in Fig.~\ref{Fig3}(b), (c), and (d),
respectively. We neglected the spin-independent energy $\hbar^2 k_z^2/2m$
for clarity. Interestingly, zero energy states emerge for both real and
imaginary parts of the spectra. Yet, the surface states (Fermi arc) are only
associated with those of zero absolute energy, which connect the Weyl
exceptional ring at the center ($\mathbf{k}_{W0}$) to those on two sides ($%
\mathbf{k}_{W\pm}$). These states are doubly degenerate eigenvectors (not
generalized ones~\cite{Tony2016,Chong2016}), one (the other) of which is
localized on the left (right) surface. Compared to the surface states
without $\gamma$, their range along $k_z$ decreases with respect to $\gamma$
because the size of the rings along the $z$ direction grows with $\gamma$.
Fig.~\ref{Fig3}(d) shows the shrinking surface states for $\gamma=0$, $0.35J$%
, $0.7J$ and $0.86J$. As $\gamma$ becomes sufficiently large, the rings
around $(k_x a_x,k_y a_y)=(0,0)$ overlap with those around $(\pi,0)$ in the $%
k_{z}$ direction and surface states completely disappear.

To measure the Weyl exceptional ring, a possible approach is to probe the
dynamics of atom numbers of each spin component after a quench~\cite%
{LuoleArxiv}. Initially, if we only keep the spin-independent optical
lattices but switch off the spin-dependent ones (contributing to the
spin-orbit coupling) and dissipation, we can load spin up atoms into the
system and the ground state is $\Psi(\mathbf{k}=0, t=0)=(%
\begin{array}{cccc}
1 & 0 & 1 & 0%
\end{array}%
)/\sqrt{2}$ since the Hamiltonian reduces to $H=-h_t\tau_x$. This state can
be driven to a state with $\mathbf{k}\neq 0$ by accelerating the optical
lattices or by applying an external gradient force. After that, the
spin-orbit coupling and dissipation can be suddenly turned on. So this state
is no longer the eigenstate of the system and the atom numbers will change
with time. For example, if $\mathbf{k}$ lies in the $b_{z-}=0$ plane, the
normalized atom number for spin down is given by
\begin{equation}
n_\downarrow \! = \frac{b_x^2+b_y^2}{4|E_\theta|^2} \Big[ %
\sum_{\lambda=\pm}e^{\lambda 2 \text{Im}(E_\theta)t/\hbar} -2\cos \Big(
\frac{2\text{Re}(E_\theta) t}{\hbar} \Big) \Big],
\end{equation}
where $n_\downarrow=N_\downarrow e^{\gamma t/\hbar}$ with $N_\downarrow$
being the atom number. Outside of the ring, ${\text{Im}}(E_\theta)=0$ and $%
n_\downarrow=(b_x^2+b_y^2)\sin^{2}(E_\theta t/\hbar)/E_\theta^2$ with an
oscillation period of $2\pi \hbar/E_\theta$ and inside the ring $\text{Re}%
(E_\theta)=0$ and $n_\downarrow=(b_x^2+b_y^2)[\sum_{\lambda=\pm}e^{\lambda2%
\text{Im}(E_\theta)t/\hbar} -2]/(4|E_\theta|^2)$ with no oscillation. The
existence of the Weyl exceptional ring will be manifested through the change
in oscillation periods. In experiments, one may choose $^{87}$Rb (bosons)
atoms and apply blue-detuned laser beams at wavelength $\lambda=767 \,$nm~%
\cite{Shuai2015} to generate the optical lattices with Weyl points. With
specific experimental settings, our model parameters are given by $%
J_{SO}=0.5J$ and $J=0.058E_{R}$, where the recoil energy is $%
E_{R}/\hbar=\hbar k_R^2/2m=2\pi \times 3.9\,$kHz with $k_R=2\pi/\lambda$ and
$\lambda$ being the wavelength of laser beams. The decay strength $\gamma$
can be experimentally tuned by controlling the intensity of the resonant
optical beam.

In summary, we have discovered a Weyl exceptional ring in a dissipative
system of Weyl points with particle gain and loss. Such a ring
is characterized by both a quantized Chern number and a quantized Berry phase,
which are defined via the Riemann surface. We further
propose an experimental scheme in cold atoms to realize the Weyl exceptional
ring, which
paves the way for future experimental investigation of such a
ring and its unusual topological properties.

\begin{acknowledgments}
We thank S. A. Yang for helpful discussions. This work was supported by the ARL, the IARPA LogiQ program, and the AFOSR MURI program.
\end{acknowledgments}

\begin{widetext}
\section{Supplemental Material}
\setcounter{equation}{0} \setcounter{figure}{0} \setcounter{table}{0} %
\renewcommand{\theequation}{S\arabic{equation}} \renewcommand{\thefigure}{S%
\arabic{figure}} \renewcommand{\bibnumfmt}[1]{[S#1]} \renewcommand{%
\citenumfont}[1]{S#1}
In the supplemental material, we will derive the semiclassical equation for
a wave packet in an optical lattice with a non-Hermitian term. The wave packet
can be constructed from Bloch wave functions:
\begin{equation}
|\Phi\rangle=\int d{\bm k}a({\bf k},t)|\phi_{n}(\bf k)\rangle,\label{eq:WavePacket}
\end{equation}
where $|\phi_{n}({\bf k})\rangle$ is a right Bloch wave function of $H_C$, the Hamiltonian describing
the system, i.e.,
$H_C |\phi_{n}({\bf k})\rangle=E_{n}({\bf k})|\phi_{n}({\bf k})\rangle$ with
$|\phi_{n}({\bf k})\rangle=e^{i{\bf k}\cdot{\bf r}}|u_{n}({\bf k})\rangle$,
$\langle u_{n}({\bf k})|u_{n}({\bf k})\rangle=1$
and $n$ being the band index; $a({\bf k},t)=|a({\bf k},t)|e^{i\gamma({\bf k},t)}$ with the amplitude $|a({\bf k},t)|$
taking the Gaussian form centered at ${\bf k}_{c}=\int d{\bf k}|a^{\prime}({\bf k},t)|^{2}{\bf k}$
with $|a^{\prime}({\bf k},t)|^{2}\equiv|a({\bf k},t)|^{2}/\int d{\bf k}^{\prime}|a({\bf k}^{\prime},t)|^{2}$
and the phase $\gamma({\bf k},t)$ incorporating a dynamical and Berry phase. Since we only consider the case
that a single band is occupied by the wave packet, we will neglect the subscript $n$ hereafter.

Using Eq.~\eqref{eq:WavePacket}, the location of the wave packet is given by
\begin{align}
{\bf r}_{c} & =\langle\Phi|\hat{{\bf r}}|\Phi\rangle/\langle\Phi|\Phi\rangle\nonumber \\
 & =-\int d{\bm k}|a^{\prime}({\bf k},t)|^{2}\partial_{{\bf k}}\gamma({\bf k},t)+\int d{\bm k}|a^{\prime}({\bf k},t)|^{2}\langle
 u({{\bf k}})|i\partial_{{\bf k}}|u({{\bf k}})\rangle\\
 & \approx -\partial_{{\bf k}_{c}}\gamma({\bf k}_{c},t)+{\bf A}({\bf k}_{c}),
\label{Eq:r}
\end{align}
where ${\bf A}({\bf k}_{c})=i\langle u({\bf k}_{c})|\partial_{{\bf k}_{c}}|u({\bf k}_{c})\rangle$ and
$\gamma({\bf k}_{c},t)=-\int_{t_{0}}^{t}dt^{\prime}\text{Re}\left[E_{{\bf k}_c(t^{\prime})}-\tilde{{\bf A}}({\bf k}_c(t^{\prime}))
\cdot\frac{d{\bf k}_c}{dt^\prime}\right]$ with $\tilde{\bf A}({\bf k}_{c})=i\langle \tilde{u}({\bf k}_{c})|\partial_{{\bf k}_{c}}|u({\bf k}_{c})\rangle$,
where the first and second terms correspond to the dynamical and Berry phases, respectively. Using Eq.~\eqref{Eq:r}, we can obtain the
semiclassical equation in terms of a velocity
\begin{equation}
{\bf v}_{c}(t_{0}) =\text{lim}_{\delta t\rightarrow 0}\frac{{\bf r}_{c}(t_{0}+\delta t)-{\bf r}_{c}(t_{0})}{\delta t}
=\partial_{{\bf k}_{c}}\bar{E}({{\bf k}_{c}})-\dot{{\bf k}}_{c}\times{\bm \Omega}({\bf k}_c)
\end{equation}
with $\bar{E}({{\bf k}_{c}})=\text{Re}[E({{\bf k}_{c}})]+\bar{{\bf A}}({\bf k}_{c})\cdot \dot{{\bf k}}_{c}$.
\end{widetext}

\end{document}